# An Automatic Cardiac Segmentation Framework based on Multi-sequence MR Image

Yashu Liu, Wei Wang, Kuanquan Wang(✉), Chengqin Ye, Gongning Luo

Harbin Institute of Technology, Harbin 150001, China
wangkq@hit.edu.cn

**Abstract.** LGE CMR is an efficient technology for detecting infarcted myocardium. An efficient and objective ventricle segmentation method in LGE can benefit the location of the infarcted myocardium. In this paper, we proposed an automatic framework for LGE image segmentation. There are just 5 labeled LGE volumes with about 15 slices of each volume. We adopted histogram match, an invariant of rotation registration method, on the other labeled modalities to achieve effective augmentation of the training data. A CNN segmentation model was trained based on the augmented training data by leave-one-out strategy. The predicted result of the model followed a connected component analysis for each class to remain the largest connected component as the final segmentation result. Our model was evaluated by the 2019 Multi-sequence Cardiac MR Segmentation Challenge. The mean testing result of 40 testing volumes on Dice score, Jaccard score, Surface distance, and Hausdorff distance is 0.8087, 0.6976, 2.8727mm, and 15.6387mm, respectively. The experiment result shows a satisfying performance of the proposed framework. Code is available at https://github.com/Suiiyu/MS-CMR2019.

**Keywords:** Ventricle segmentation, Histogram match, LGE-CMR, Data augmentation.

## 1 Introduction

Cardiac MRI is a significant technology for cardiac function analysis. Benefiting from this technology, the doctor can evaluate the heart function noninvasively. There are many kinds of Cardiac MRI modalities, such as balanced-Steady State Free Precession (b-SFFP) and LGE. b-SSFP can learn the cardiac motions and obtain a clear boundary of cardiac. LGE CMR can enhance the infarcted myocardium, appearing with distinctive brightness compared with the healthy tissues. LGE CMR is widely used to study the presence, location, and extent of myocardium infarction (MI) in clinical studies [1, 2]. Exactly extracting the ventricles and myocardium from LGE is crucial for MI therapy. However, the infarcted myocardium is enhanced, meanwhile, the healthy myocardium is suppressed. Hence, the boundaries of the ventricles and myocardium are bedimmed on the LGE CMR.

In the clinical application, ventricle segmentation on LGE CMR image still relies on manual segmentation. However, manual segmentation is tedious and subjective. The automatic segmentation method is more efficient and objective. Kurzendorfer et al. [3] proposed an automatic framework to segment left ventricle (LV). They firstly initialized the LV by a two-step registration method and then adopted principal components to estimate the LV. At last, the myocardium was refined on the poly space. Oktay et al. [4] incorporated global shape information into CNN. They utilized auto-encoder to estimate the global shape information of LV and then it was adopted to constrain the segmentation model. Duan et al. [5] proposed a combined CNN and level set model to segment ventricles. The probability maps of ventricles and myocardium are estimated by CNN. Then they initialized the energy function of the level set by the probability maps. Khened et al. [6] adopted a densely connected CNN model with inception block to segment 2D cardiac MRI. There are also other researchers interesting on the ventricles, myocardium and other tissues MR segmentation [7-13]. Their methods are mostly based on CNN. Besides, these methods rely on a large number of training data. However, in our situation, there are just 5 labeled and 40 unlabeled LGE CMR with about 15 slices of each volume. The rare data cannot guarantee training an efficient cardiac segmentation model from scratch. Although the registration method, such as atlas, is often utilized on the rare data segmentation, it has some shortages. In order to obtain the label for the unlabeled data, the atlas set must be labeled data. Moreover, it will deform the original data and decreases the data diversity. Hence, we utilize histogram match technology to achieve effective augmentation based on b-SSFP modality CMR data, which has 35 labeled volumes, to solve the lack of data problem. Histogram matching [14] technology is efficient and does not deform the shape of the original data. Hence, we can adopt other modalities data while the data diversity is maintained. Then, we adopt this augmented dataset to train a cardiac segmentation model. At last, we utilize a label-vote strategy and connected component analysis to get the final segmentation.

The rest of this manuscript is organized as follows: we introduce our method in Section 2. Results are analyzed in Section 3. Finally, we conclude this manuscript in Section 4.

## 2    Method

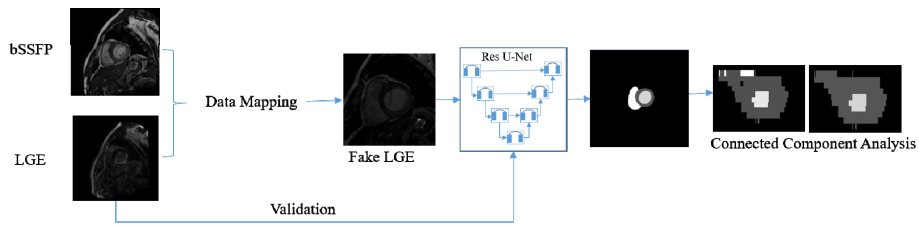

**Fig. 1.** The proposed framework for ventricles and myocardium segmentation on LGE CMR image. The white, light gray and dark gray correspond to represent right ventricle, left ventricle and left ventricle myocardium.

The whole structure of the proposed framework is shown in Fig. 1. There are three steps. Firstly, we pre-process the volumes into images and then we map the b-SSFP images on the LGE images to generate fake LGE images. Secondly, the fake images are fed into the Res-UNet [15, 16] model. Our model is trained based on the leave-one-out strategy. The final prediction is determined by all models. Thirdly, the predicted results are reconstructed to the original shape and a connected component analysis is adopted to keep the maximum component for each class as the final segmentation result.

### 2.1 Data Processing

The dataset is coming from the 2019 Multi-sequence Cardiac MR Segmentation Challenge (MS-CMR2019)[1]. It published 45 patients CMR data with three modalities, T2, b-SSFP, and LGE. There are 35 labeled T2 CMR data with about 3 slices of each patient, and 35 labeled b-SSFP CMR data with about 11 slices of each patient, and just 5 labeled LGE CMR data with about 15 slices of each patient. The rest volumes are un-labeled data. The main purpose of this challenge is segmenting left ventricle (LV), right ventricle (RV) and left ventricle myocardium (LVM) from LGE CMR data. The rarely labeled target data increases the challenge sharply. To enlarge the number of labeled LGE CMR data, we utilize histogram match on the other labeled modalities.

According to the data analysis, the b-SSFP data has the similarity slices with LGE data of each patient and it has a clearer boundary than T2 modality data. Considering the data-matching problem and data quality, we just utilize b-SFFP data to assist the cardiac segmentation on LGE data. We find that the main difference between LGE image and b-SSFP image is the appearance. The shape of the heart among the same patient is similar. Hence, we utilize histogram match to generate the fake LGE data. Histogram match is an easy and efficient data pre-process for this challenge. It matches the histogram of the source image to the target histogram by establishing the relationship between the source image and the target image. Moreover, the shape of the source image is still maintained. That is mean that the label of fake LGE CMR images is still consistent with the original b-SSFP CMR images.

In order to retain the data diversity, each b-SSFP image has its own target LGE image histogram. Because of the original b-SSFP data and LGE data have different data scope. The scope of the short axis of LGE data is about twice larger than b-SSFP's. Hence, we resize the LGE data into the shape of b-SSFP data. Then, we obtain 2D images of the short axis from the resized data. So far, we have got the consistent image size and number of b-SSFP and LGE. The target histogram for each b-SSFP image is calculating from the corresponding LGE image. Fig. 2 presents an example of a resized LGE image, b-SSFP image, and fake LGE image. a, b are corresponding to the short axis of LGE and b-SSFP; d, e are corresponding to the long axis of LGE and b-SSFP; c, f are the short axis and long axis of fake LGE which are generating from real LGE and b-SSFP. Image c and f owns the shape information of b-SSFP image and the appearance information of LGE image.

---

[1] https://zmiclab.github.io/mscmrseg19/

Our model is trained on 2D images, which are extracting from fake LGE data and real original labeled LGE data. In order to keep the same input to the model, we resize all images into (256, 256). After data analysis, we center crop the resized images into (144,144) to filter the unrelated background. The output of the model will do the inverse operation to keep the data consistency. Moreover, the evaluation is performed on the 3D volumes.

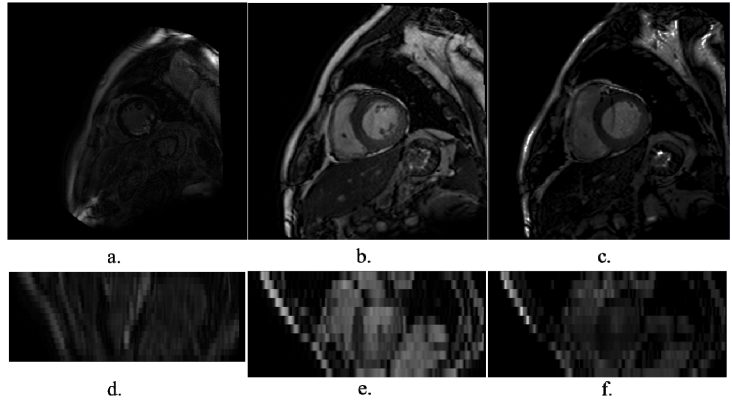

**Fig. 2.** The example of LGE, b-SSFP, and fake LGE. a-c: Corresponding to the short axis LGE, b-SSFP, and fake LGE generating from a and b; d-f: Corresponding to the long axis LGE, b-SSFP, and fake LGE.

### 2.2 Implementation

The segmentation model is a Res-UNet, which utilizes residual connection on the convolutional block. Each convolutional block contains two 3*3 convolutional layers with ReLU activation function and batch normalization. We adopt 4 down-sampling blocks as the encoder and corresponding up-sampling blocks as the decoder. The last block utilizes a dropout layer with 0.5 drop rate to overcome the over-fitting problem. The output layer is a 1*1 convolutional layer with a Softmax activate function. The model is implemented using Keras based on NVIDIA 2080 Ti GPU.

In order to maximize the data utilization, we divide the 5 labeled LGE volumes into 5 groups by the leave-one-out strategy. At last, we have trained 5 models, and the training data of each model consists of 35 fake LGE volumes and 4 real LGE volumes. The rest one real LGE volume is utilized to evaluate the model. The final prediction is determined by the average of these models. Each model has trained 300 epochs with 0.001 learning rate and 8 batch size. The training time is about 1 hour for each model. Moreover, we utilize a weighted cross entropy loss function to solve the class imbalance problem. The loss function is shown in Eq. 1:

$$wCE = -\sum_{c=0}^{4} w_c (\sum_{i=1}^{N} g_{c_i} \log p_{c_i}),$$

$$w_c = \frac{\sum g_c}{\sum g}$$

(1)

where $c$ is the class index; $i$ is the pixel index; $g_{c_i}$ and $p_{c_i}$ represent the ground truth class and prediction class of pixel $i$. The weight $w_c$ is calculated by the ratio of each class in the all labeled set. And $g$ is the all labeled pixels set.

After training the segmentation model, we reconstruct the prediction results in the original shape. Then, a connected component analysis is performed to remain the largest connected region for each class as the final segmentation result. Our segmentation model is evaluated by the official evaluation metrics, which are Dice score, Jaccard score, Surface distance, and Hausdorff distance. Dice score and Jaccard score are overlapped metrics. They evaluate the overlap ratio between the ground truth and predicted result. However, they have a shortage on the boundary details of the subject. Although similarity metrics, Surface distance and Hausdorff distance, are mainly focused on the similarity between the ground truth and predicted result, they are sensitive on the noisy. Utilizing both of these metrics can complement one another perfectly. Hence, the segmentation model can be over-all evaluated. Notice that the Dice score is the main metric.

## 3 Experimental Results

The score of metrics during the validation stage is shown in Table.1. These scores are the mean value of the three classes, which are calculated by average operation without weighted. The segmentation model has a satisfying performance on the overlap metrics. Due to the model is trained on the short axis, the performance on the similarity metrics are worse than overlap metrics. Fig. 3 represents the segmentation results and corresponding ground truths of patient 1 and patient 2. The green and red contours represent ground truth and segmentation result, respectively. We select three representative slices to show the result. The result shows that our prediction contours can perfectly fit the ground truth.

**Table 1.** Segmentation results of the validation stage. SD and HD correspond to the abbreviation of Surface distance and Hausdorff distance.

| Patient | Dice Score | Jaccard Score | SD (mm) | HD (mm) |
|---------|------------|---------------|---------|---------|
| #1      | 0.9289     | 0.8685        | 0.3873  | 6.6570  |
| #2      | 0.9461     | 0.8997        | 0.3012  | 14.2289 |
| #3      | 0.9277     | 0.8665        | 0.3761  | 5.8568  |
| #4      | 0.9416     | 0.8899        | 0.2801  | 4.8050  |
| #5      | 0.9128     | 0.8439        | 0.4608  | 5.7329  |
| Mean    | 0.9315     | 0.8737        | 0.3611  | 7.4561  |

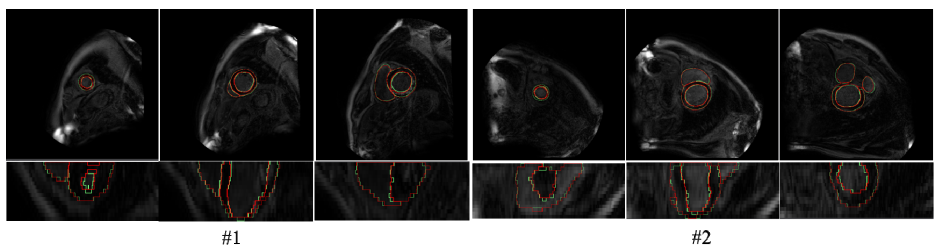

**Fig. 3.** Segmentation result and ground truth of patient 1 and patient 2. The first row is the short axis view; the second row is the long axis view. Color representation: green-ground truth; red-segmentation result.

Fig. 4 exhibits the metrics of 40 patients during the testing stage. The testing segmentation result is evaluated by the organizer. The patient IDs are anonymous, but their orders are consistent across the four metrics. We obtain a satisfying result on the testing set except for three worse results, 5th, 19th, and 39th. From these sub-pictures, we can find that the LV cavity has the regular shape and largest area in the three classes. It gets the highest scores across all metrics. On the opposite, the model performs worse on the irregular RV.

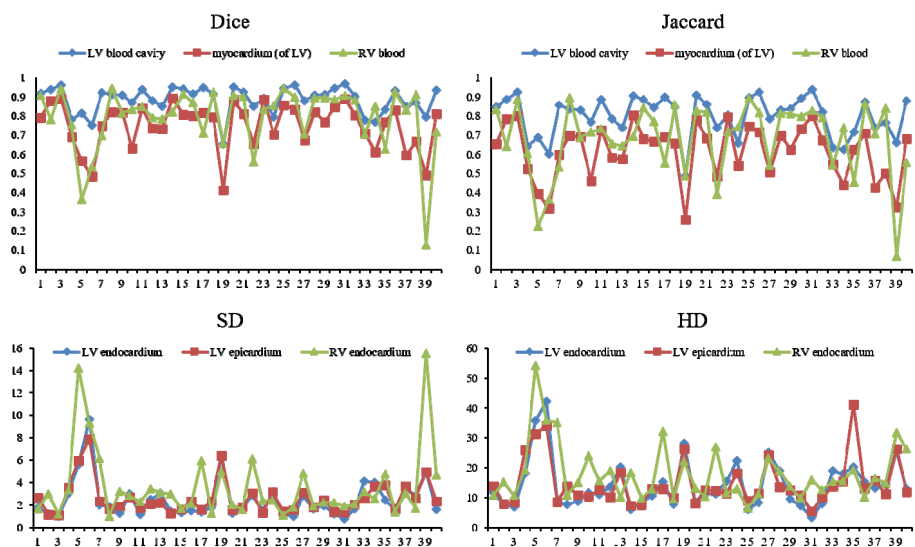

**Fig.4.** The metrics of 40 testing volumes. The patient IDs are anonymous, but their orders are consistent across the four metrics. SD and HD correspond to the abbreviation of Surface distance and Hausdorff distance.

Fig. 5 exhibits the segmentation results of patient 6 and patient 24, which are randomly chosen from the testing dataset. The green, red and yellow contours represent LVM, RV and LV, respectively. The three columns of each patient are from three different slices in order to demonstrate a comprehensive result of the proposed model.

The model obtains a perfect performance on the short axis, especially the LV. However, there still some shortages on the long axis view due to our segmentation model is processed on the short axis.

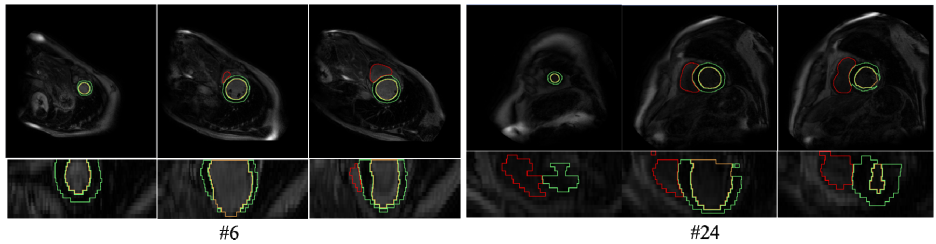

**Fig.5.** Segmentation result of patient 6 and patient 24. The first row is the short axis view; the second row is the long axis view. The three columns of each patient are from three different slices according to the long axis. Color representation: green-left ventricle myocardium; yellow-left ventricle; red-right ventricle.

## 4   Conclusion

LGE CMR is an efficient technology to identify infarcted myocardium. In this paper, we proposed an automatic framework for LGE CMR segmentation. This framework contains three steps. Firstly, we adopted a histogram match process on the b-SSFP images to generate fake LGE images. Secondly, we divided the labeled LGE images into 5 groups through the leave-one-out strategy. Our segmentation model, Res-UNet, was trained based on the fake LGE images and labeled LGE images. Thirdly, the final prediction of the model was reconstructed and a connected component analysis process was done on these data to keep the maximum connected component for each class as the final segmentation. The final segmentation is evaluated by the organizer, and the mean metrics score of Dice score, Jaccard score, Surface distance, and Hausdorff distance are corresponding to 0.8087, 0.6976, 2.8727mm, and 15.6387mm. There are three worse volumes out of 40 testing volumes. The performance on the most volumes are satisfied.

**Acknowledgment.** This work was supported by the National Key R&D Program of China under Grant 2017YFC0113000.